\documentclass[twoside]{dis07}
\usepackage[latin1]{inputenc}
\usepackage[dvips]{graphicx,epsfig,color}
\usepackage{wrapfig,rotating}
\usepackage{amssymb,amsmath,array}

\def\lesssim{\ \hbox{\raise 2pt \hbox{$<$} \kern -13pt
                     \lower 3pt \hbox{$\sim$}}\ }
\def\greatersim{\ \hbox{\raise 2pt \hbox{$>$} \kern -13pt
                     \lower 3pt \hbox{$\sim$}}\ }

\begin{document}
\begin{flushright}
CERN-PH-TH/2007-098\\
\end{flushright}
\vskip 1.5 cm 
\begin{center}
{\Large \bf \boldmath Coordinate-space picture and $x \to 1$   
singularities \vskip 0.1 cm 
at fixed $k_\perp$}
\end{center}
\vskip 0.5 cm 
\begin{center}
{F.~Hautmann}
\end{center}
\vskip 0.2 cm 
\begin{center}
{CERN, PH-TH Division,
Geneva, Switzerland and\\
Institut f{\" u}r Theoretische Physik,
Universit{\" a}t Regensburg, Germany}
\end{center}
\vskip 1.5 cm 
\begin{center}
{Abstract}
\end{center}
\vskip 0.2 cm 
\noindent We discuss    
 ongoing progress towards precise 
characterizations of parton distributions at fixed transverse momentum, 
focusing on   matrix elements in coordinate space 
and the  treatment of   endpoint singularities. 
\vskip 2.5 cm 
{\em \hspace*{0.2 cm} Based on talks 
 given at the {\rm HERA and the LHC}   Workshop (Hamburg, March 2007) \\
\hspace*{2.9 cm}    and at the  {\rm DIS07}   Workshop (Munich, April 2007).
}

\title{Coordinate-space picture and $x \to 1$   
singularities\\ at fixed $k_\perp$ 
}

\author{F.~Hautmann
%
%
\vspace{.3cm}\\
%
CERN, PH-TH Division,
Geneva, Switzerland and\\
Institut f{\" u}r Theoretische Physik,
Universit{\" a}t Regensburg, Germany
}

\maketitle

\begin{abstract}
We discuss    
 ongoing progress towards precise 
characterizations of parton distributions at fixed transverse momentum, 
focusing on   matrix elements in coordinate space 
and the  treatment of   endpoint singularities. 
\end{abstract}

Parton distributions unintegrated in transverse momentum are 
naturally defined for small x 
via high-energy factorization~\cite{hef}. 
This relates  
off-shell matrix elements with  physical cross sections 
at x  $\to 0$,  
and  gives a  well-prescribed  method to introduce 
 unintegrated parton distributions  
  in a gauge-invariant manner. 

The question of how 
to characterize gauge-invariantly a $k_\perp$ distribution 
  over the whole phase space, on the other hand,  
 is more difficult and not yet fully answered.  
Its relevance was already emphasized long ago in the context of 
 Sudakov processes~\cite{collsud}, 
  jet physics~\cite{cs81}, 
exclusive production~\cite{brodlep}, 
spin physics~\cite{mulders}. 
Although a complete framework is still missing,  
much work is currently underway on this subject, see 
e.g.~\cite{qiuvogel,collinsqiu,ceccopieri,goeke,fhfeb07,idimeh,ji06}. 
The discussion that follows focuses on aspects  related 
to the gauge-invariant operator matrix elements and regularization 
methods for lightcone divergences.

\begin{figure}[htb]
\vspace{50mm}
\includegraphics{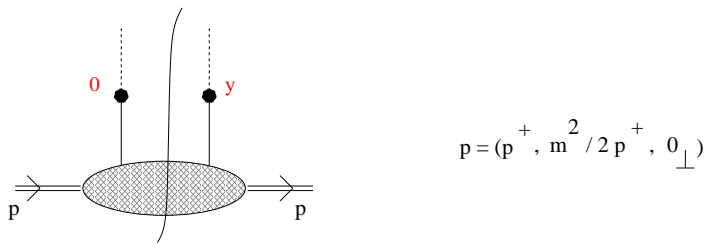}
\caption{Quark distribution function in the target of momentum p.}
\label{fig:pdf}
\end{figure}

To ensure gauge invariance, the  approach commonly used   is to 
 generalize     
 the  matrix elements  that serve to define 
 ordinary parton distributions 
to the case 
of field operators  at non-lightcone distances~\cite{mulders,beli}.  
This leads one to consider the matrix element for the quark distribution 
  (Fig.~\ref{fig:pdf})  

\begin{equation}
\label{coomatrel}
  {\widetilde f} ( y  ) =
  \langle P |  {\overline \psi} (y  )
  V_y^\dagger ( n ) \gamma^+ V_0 ( n )
 \psi ( 0  ) |
  P \rangle  \hspace*{0.3 cm}  
\end{equation}
with the  quark fields $\psi$ evaluated at distance 
$y = ( 0 , y^- , y_\perp  )$ for arbitrary $y^-$ and $y_\perp$,
 and  the eikonal-line operators $V$ given by 
\begin{equation}
\label{defofV}
V_y ( n ) = {\cal P} \exp \left(  i g_s \int_0^\infty d \tau \
n^\mu A_\mu (y + \tau \ n) \right)
    \hspace*{0.3 cm} ,   
\end{equation}
where $n$ is the direction of the eikonal line and $A$
is the gauge field.

However, while the use of Eq.~(\ref{coomatrel}) does not pose major 
problems at tree level, 
it  becomes more subtle at the level of radiative corrections. 
Part of the subtleties are associated with  incomplete KLN cancellations that come 
 from  measuring $k_\perp$ in the 
initial state~\cite{collsud,jccrev03}. These may appear as uncancelled 
divergences near the endpoints for certain lightcone momentum 
components~\cite{brodsky01}. Another set of issues are   
  associated with the integration  
over all transverse momenta, and involve    the relation  of 
unintegrated parton distributions 
with the ordinary ones~\cite{cch93,jcczu,watt} and 
the treatment of ultraviolet divergences. 
As observed  in~\cite{jccfh00}  for the case of the Sudakov 
form factor, the choice of a particular 
regularization method for the lightcone divergences   also 
affects   integrated distributions and  ultraviolet subtractions.

 In~\cite{fhfeb07}  
these effects are examined by an explicit calculation at one loop 
using techniques for the expansion of nonlocal operators. 
The answer for the coordinate-space  matrix element is 
analyzed 
in powers of  $y^2$, separating logarithmic contributions 
from long distances and short distances, 
\begin{eqnarray}
\label{EbmEaexpand}
 {\widetilde f}_{1} (y)  &=&   
{{ \alpha_s C_F     } \over { \pi  } } 
\ p^+
\int_0^1 dv  \ { v \over { 1 - v }} \ 
\left\{ 
\left[ e^{ i p \cdot y v}  - e^{ i p \cdot y } \right] 
\ \Gamma ( 2 - { d \over 2} ) \ 
( { {4 \pi \mu^2} \over \rho^2} )^{2-d/2}
\right. 
\nonumber\\
 &+& \left.
e^{ i p \cdot y v} \ \pi^{2-d/2} \ 
\Gamma (  { d \over 2} - 2 ) \ 
(- y^2  \mu^2)^{2 - d/2}
+ \cdots 
\right\} \hspace*{0.2 cm}  , 
\end{eqnarray}
where $\mu$ is the  dimensional-regularization scale 
  and $\rho$ is an infrared  mass regulator.  
The lightcone singularity $v \to 1$ 
corresponds to the exclusive phase-space boundary $x = 1$.  
  The singularity cancels for ordinary parton distributions  
 (first term in the right hand side of Eq.~(\ref{EbmEaexpand})) 
but it    
 is present, even at $d \neq 4$ and finite 
$\rho$,   in subsequent terms,  which contribute to the unintegrated parton 
distribution~\cite{fhfeb07}. This is then treated on the same footing as a 
physical correlation function, to be expanded in terms of the ordinary parton 
distributions with nontrivial, perturbatively 
calculable coefficient functions~\cite{cch93,jcczu}. 

\begin{figure}[htb]
\vspace{35mm}
\includegraphics{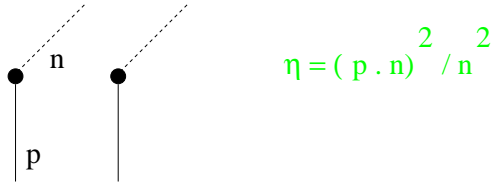}
\caption{Cut-off regularization for the quark matrix element.}
\label{fig:cutoff}
\end{figure}

Traditionally the effect of endpoint 
 singularities is suppressed by  the use of  
a cut-off. 
This is likely   the case, for instance,  in 
 existing  Monte-Carlo event 
generators that implement  
unintegrated  parton 
distributions~\cite{marchweb92,lonn,junghgs,golec-mc,krauss-bfkl}. 
A cut-off is also implemented in   
 treatments~\cite{collsud,ji06,jiyuan} 
based on regularizing the parton-distribution matrix element  
by taking the eikonal line $n$ to be 
non-lightlike (Fig.~\ref{fig:cutoff}),  combined with 
evolution equations in the cut-off 
parameter $\eta = (p \cdot n)^2 / n^2$~\cite{cs81,korchangle}.  
 One-loop formulas in coordinate space corresponding  to the 
regularization method of Fig.~\ref{fig:cutoff} are given  
in~\cite{fhfeb07}.  
This method leads to a cut-off 
in $x$ at fixed  $k_\perp$ of order 
\begin{equation}
\label{mink}
1 - x
\greatersim
k_\perp / \sqrt{4 \eta}
    \hspace*{0.3 cm} .  
\end{equation}

However, cut-off regularization is not very  
well-suited for applications 
 beyond the leading order. 
Furthermore, as the two lightcone limits $y^2 \to 0$ and $n^2 \to 0$ do not commute, 
a residual dependence on  the regularization 
parameter $\eta$ is left  after integrating in $k_\perp$ 
the distribution defined with the cut-off. The relation with the standard 
operator product expansion is therefore not so transparent.

\begin{figure}[htb]
\vspace{65mm}
\includegraphics{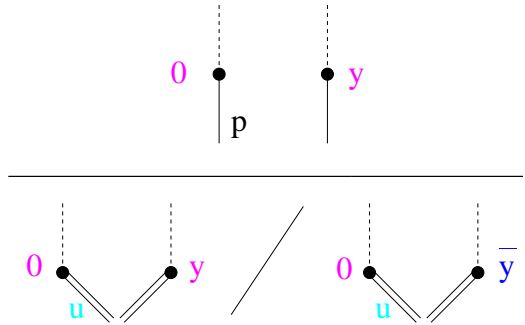}
\caption{Matrix element with subtractive regularization.}
\label{fig:subtr}
\end{figure}

An alternative approach  is based on the subtractive regularization 
method~\cite{jccfh00,jccfh01}. As explained 
in~\cite{jccrev03}, in this approach  the eikonal 
$n$ is kept in lightlike direction but the singularities 
are canceled by multiplicative, gauge-invariant factors 
given by eikonal-line  vacuum expectation  
values. 
The matrix element with subtraction factors  is pictured in 
 Fig.~\ref{fig:subtr}, 
where ${\bar y} = ( 0 , y^- , 0_\perp  )$, and $u$ is 
the direction of an auxiliary (non-lightlike) eikonal that provides 
a gauge-invariant regulator near $x = 1$ and  cancels in 
the matrix element at $y_\perp = 0$~\cite{fhfeb07}.  
The form of the counterterms is simple in coordinate space, 
where it can be given in terms of compact 
all-order expressions.

The subtractive method is more systematic than the cut-off, and likely more 
suitable for using unintegrated parton distributions at subleading-log 
level. 
It can  be useful  for incorporating  
the unintegrated formulation in 
   parton shower approaches~\cite{jcczu,jccfh01,bauermc}.  Also, subleading 
accuracy is needed for matching large-x contributions   
  with  calculations at small x~\cite{watt,krauss-bfkl,jungkoti} and  
 in the  Sudakov region~\cite{cpyuan,cpyuan1}. 

The techniques discussed above 
will  be instrumental 
  to analyze factorization and evolution for 
$k_\perp$ parton distributions 
 with increased 
precision~\cite{collinsqiu,ceccopieri}.  
The issue of soft gluon exchanges 
with spectator partons is revisited in~\cite{collinsqiu} 
for hard pp collisions. A 
potential breakdown of  factorization  
 at  high order of perturbation theory is discussed 
(N$^3$LO correction to dihadron production, with two soft and 
one collinear partons), which  would be of  interest to verify by 
calculation. Also, it will be interesting to investigate  how 
the argument of~\cite{collinsqiu} is modified if one takes account 
of  destructive-interference effects due to soft gluon coherence.  
  A better understanding of these issues will help 
 improve  the  present  accuracy 
 in current phenomenological studies of  the effects of 
partons' transverse momentum~\cite{krauss-bfkl,jungkoti,cpyuan}.


\begin{footnotesize}




\begin{thebibliography}{99}

\bibitem{hef} 
     S.~Catani, M.~Ciafaloni and F.~Hautmann, Phys. Lett. 
     {\bf B242}  (1990) 97; Nucl.\ Phys.\ {\bf B366} (1991) 135. 
\bibitem{collsud}
    J.C.\ Collins, in {\em Perturbative Quantum Chromodynamics},
    ed.~A.H.~Mueller, World Scientific 1989, p.~573. 
\bibitem{cs81}
    J.C.~Collins and D.E.~Soper,  Nucl.\ Phys.\ {\bf B193} (1981)  381. 
\bibitem{brodlep}
    S.J.~Brodsky and G.P.~Lepage, in {\em Perturbative 
    Quantum Chromodynamics},
    ed.~A.H.~Mueller, World Scientific 1989, p.~93. 
\bibitem{mulders}
    P.~Mulders and    R.D.~Tangerman,  Nucl.\ Phys.\ {\bf B461}  (1996) 197; 
     D.~Boer and P.~Mulders, Phys.\ Rev.\ {\bf D} 57 (1998) 5780.
\bibitem{qiuvogel}
   J.W.~Qiu, W.~Vogelsang and F.~Yuan, arXiv:0706.1196;   
   X.~Ji, J.W.~Qiu, W.~Vogelsang and F.~Yuan, Phys.\ Rev.\ Lett.\ {\bf 97} 
   (2006) 082002.
\bibitem{collinsqiu}
     J.C.~Collins and J.W.~Qiu, arXiv:0705.2141. 

\bibitem{ceccopieri}
     F.~Ceccopieri and L.~Trentadue, arXiv:0706.4242. 

 
\bibitem{goeke}
     K.~Goeke, S.~Meissner and A.~Metz, hep-ph/0703176.  
\bibitem{fhfeb07}
     F.~Hautmann, hep-ph/0702196.

\bibitem{idimeh}
     A.~Idilbi and T.~Mehen, hep-ph/0702022. 
\bibitem{ji06}
     P.~Chen, A.~Idilbi and X.~Ji,
     Nucl.\ Phys.\ {\bf B763} (2007) 183.


\bibitem{beli}
     A.V.~Belitsky,  X.~Ji and    F.~Yuan, 
     Nucl.\ Phys.\ {\bf B656}  (2003) 165. 

\bibitem{jccrev03}
    J.C.~Collins,
    Acta Phys.\ Polon.\ B {\bf 34} (2003) 3103.
\bibitem{brodsky01}
     S.J.~Brodsky, D.S.~Hwang, B.Q.~Ma and  I.~Schmidt,
     Nucl.\ Phys.\ {\bf B593} (2001) 311.
\bibitem{cch93}
      S.~Catani, M.~Ciafaloni and F.~Hautmann, 
     Phys. Lett.      {\bf B307}  (1993) 147; S.~Catani and F.~Hautmann,
         Nucl.\ Phys.\ {\bf B427} (1994) 475. 
\bibitem{jcczu}
     J.C.~Collins and X.~Zu, JHEP {\bf 0503} (2005) 059.
\bibitem{watt}
     A.D.~Martin,  M.G.~Ryskin and G.~Watt, Eur.\ Phys.\ J.\  C{\bf 31} (2003) 73. 
\bibitem{jccfh00}
    J.C.~Collins and F.~Hautmann,
    Phys.\ Lett.\ B {\bf 472} (2000) 129. 

\bibitem{marchweb92}
    G.~Marchesini and B.R.~Webber,
    Nucl.\ Phys.\ {\bf B386} (1992) 215.
\bibitem{lonn}
    G.~Gustafson, L.~L{\" o}nnblad and G.~Miu,
    JHEP {\bf 0209} (2002) 005;
    L.~L{\" o}nnblad and M.~Sj{\" o}dahl,
    JHEP {\bf 0402} (2004) 042,   JHEP {\bf 0505} (2005) 038.
\bibitem{junghgs}
    H.~Jung, Mod.\ Phys.\ Lett.\ A19  (2004) 1.
\bibitem{golec-mc}
    K.~Golec-Biernat, S.~Jadach, W.~Placzek, P.~Stephens and  M.~Skrzypek, 
     hep-ph/0703317. 

\bibitem{krauss-bfkl}
    S.~H{\" o}che, F.~Krauss and T.~Teubner, arXiv:0705.4577. 

\bibitem{jiyuan}
    X.~Ji, J.~Ma and F.~Yuan, Phys.\ Rev.\ D {\bf 71}
    (2005) 034005, 
    JHEP {\bf 0507} (2005) 020.
\bibitem{korchangle}
     G.P.~Korchemsky and  G.~Marchesini,
     Phys.\ Lett.\ B {\bf 313} (1993) 433; 
     G.P.~Korchemsky,
     Phys.\ Lett.\ B {\bf 220} (1989) 62. 
\bibitem{jccfh01}
    J.C.~Collins and F.~Hautmann,
    JHEP {\bf 0103} (2001) 016.

\bibitem{bauermc}
     C.W.~Bauer and M.D.~Schwartz, hep-ph/0607296.

\bibitem{jungkoti}
     H.~Jung, A.V.~Kotikov, A.V.~Lipatov and N.P.~Zotov, 
     arXiv:0706.3793. 

\bibitem{cpyuan}
      C.P.~Yuan, talk at DIS07 Workshop, Munich, April 2007. 

\bibitem{cpyuan1} 
      P.M.~Nadolsky, N.~Kidonakis, F.I.~Olness and C.P.~Yuan, 
      Phys.\ Rev.\  D {\bf 67}  (2003) 074015; 
      R.~Brock, F.~Landry, P.M.~Nadolsky and C.P.~Yuan, 
      Phys.\ Rev.\  D {\bf 67}  (2003) 073016. 






\end{thebibliography}
%

\end{footnotesize}


\end{document}